# Evaluación de las propiedades tribológicas de materiales compuestos de matriz metálica (MMCs) procesados por técnicas de fabricación aditiva con haz láser (SLM)

# Evaluation of the tribological properties of metal matrix composite materials (MMCs) obtained by additive manufacturing using selective laser melting (SLM)

E. Martínez[1], O. A. González-Estrada[2], A. Martínez[3]

[1]Instituto Tecnológico Metalmecánico, Mueble, Madera, Embalaje y Afines, AIDIMME, Valencia, España.
Email: elkin.martinez@aidimme.es
[2]School of Mechanical Engineering, Universidad Industrial de Santander, Ciudad Universitaria, Bucaramanga, Colombia.
Orcid: 0000-0002- 2778-3389, Email: agonzale@uis.edu.co
[3]Grupo Nuevas Tecnologías, Universidad de Santander, Bucaramanga, Colombia, Email: alejandrom@udes.edu.co

**RESUMEN**

En este artículo se investigaron las propiedades mecánicas y tribológicas de materiales compuestos de matriz de acero (acero inoxidable 316L reforzado con partículas cerámicas $Cr_3C_2$) procesados por tecnologías de fabricación aditiva de fusión selectiva con láser (SLM). Se estudió el comportamiento a desgaste con el ensayo "pin-on-disk" a temperatura ambiente y se observó la superficie desgastada mediante microscopía electrónica de barrido (SEM). Los resultados indicaron que el coeficiente de fricción no tiene una tendencia clara o relación directa cuando se varía el porcentaje de refuerzo mientras que la tasa de desgaste disminuye con el aumento del contenido de refuerzo. Las mejores propiedades se obtuvieron con un 6% en peso de refuerzo.

**PALABRAS CLAVE:** Ensayo "pin-on-disk", Coeficiente de fricción, Tasa de desgaste, Materiales compuestos de matriz metálica (MMCs), Fabricación aditiva (AM), Fusión selectiva con láser (SLM), Superficie de desgaste modelo.

**ABSTRACT**

The mechanical and tribological properties of steel matrix composites (316L stainless steel reinforced with $Cr_3C_2$ ceramic particles) were investigated. The steel matrix composites (SMC) with three reinforcing percentages (3, 6, and 9 wt.%) were manufactured by Additive Manufacturing technologies (SLM: Selective Laser Melting). The wear behaviour was studied by using a pin-on-disk wear test at room temperature. The worn surface was analysed using Scanning Electron Microscopy (SEM). The results indicated that the friction coefficient does not have a clear tendency or direct correlation with the reinforcement variation while the wear rate decreases with increasing content of reinforcement. The better properties combination were achieved with 6 wt.% of reinforcement.

**KEYWORDS:** Pin-on-disk test, Friction coefficient, Wear rate, Metal matrix composites (MMCs), Additive Manufacturing (AM), Selective Laser Melting (SLM), Worn surface.

## 1. INTRODUCCIÓN

La tecnología de fabricación aditiva *Selective Laser Melting* (SLM), enmarcada dentro de la clasificación de fusión en lecho de polvo (*powder bed fusion*), hace que sea posible la obtención de piezas complejas mediante la sucesiva deposición de material, capa a capa, y posterior fusión selectiva con el haz láser, directamente a partir de un fichero digital con la geometría de la pieza CAD 3D [1, 2]. También se cree que este proceso disminuye el tiempo de producción de piezas complejas, maximiza el uso del material y se considera medioambientalmente compatible [3].

Los estudios que se están realizando con la tecnología SLM se centran en el desarrollo de parámetros de proceso, microestructura, corrosión, biocompatibilidad y evaluación de propiedades mecánicas, en materiales metálicos tales como aceros inoxidables y aceros de herramienta [4-7], aleaciones de aluminio del tipo Al-12Si y AlSi10Mg [8-10], titanio C.P., Ti6Al4V y aleaciones de titanio beta [11-14], aleaciones base níquel (como nitinol, inconeles y waspaloys) para propiedades de alta temperatura [15-18], aleaciones base Co (CoCrMo, CoCrMoW) para implantes dentales [19, 20], cobre comercialmente puro y aleaciones de cobre [21, 22]. En conjunto, la mayoría de las investigaciones muestran que las piezas producidas por SLM tienen mejores propiedades que las producidas por técnicas convencionales, tal como la fundición [23].

En la actualidad, se han hecho algunos intentos para mejorar aún más las propiedades (mecánicas y tribológicas) de las piezas procesadas por SLM mediante la adición de partículas de refuerzo al material metálico, es decir, la obtención de materiales compuestos de matriz metálica (MMCs). Por ejemplo, Song et al. [24], añadieron partículas de SiC (2.2% en peso) al polvo de Fe puro y lo procesaron por SLM, obteniendo un aumento de casi dos veces la resistencia última del Fe sin reforzar (357 a 764 MPa). Del mismo modo, L. Hao et al. [25], fabricaron con SLM; materiales compuestos de matriz de acero reforzado con hidroxiapatita (5 vol.%) y encontraron que la microdureza Vickers del compuesto 316L+5%vol. de HA y del 316L son 241.4 y 212.1 HV, respectivamente. Esta mayor dureza del compuesto puede atribuirse a un tamaño de grano más fino. Attar et al. [26], han producido materiales compuestos de Ti-TiB con tres niveles de porosidad diferentes (10%, 17% y 37%) por fusión selectiva por láser (SLM). Las propiedades mecánicas de estas muestras porosas disminuyen con el aumento del nivel de porosidad. El límite elástico y el módulo de elasticidad de las piezas porosas Ti-CP oscilan entre 113-350 MPa y 13-68 GPa, respectivamente, que son mucho más bajos que los de Ti-TiB porosos (234-767 MPa y 25-84 GPa respectivamente), debido principalmente al efecto endurecedor inducido por las partículas de TiB en las muestras Ti-TiB. También hay estudios que adicionan partículas de TiC a una matriz de Ti-CP [27] y a una matriz de aluminio (AlSi10Mg) [28] para la producción de materiales compuestos utilizando SLM. En ambos casos, se obtuvo un bajo coeficiente de fricción y tasa de desgaste en comparación con los materiales base sin refuerzo. Dadbakhsh et al. [29] han demostrado el uso de $Fe_2O_3$ como refuerzo para el aluminio puro. A pesar de la influencia negativa sobre la densidad, la dureza aumenta significativamente con el incremento del contenido de $Fe_2O_3$ debido a la mejora de las características microestructurales de la matriz reforzada con partículas.

En las investigaciones existentes del procesado de materiales compuestos con la tecnología SLM se puede observar que la mayoría de los refuerzos utilizados son de naturaleza cerámica (SiC, TiC, $Fe_2O_3$, etc.), que en la mayoría de los casos hay una reacción entre el refuerzo y la matriz durante el procesado por SLM y que se obtiene una mejora de las propiedades mecánicas (resistencia a costa de ductilidad) y tribológicas. El presente trabajo se centra en el estudio de las propiedades tribológicas de MMCs procesados por tecnologías de fabricación aditiva con haz láser.

## 2. DESARROLLO EXPERIMENTAL

### 2.1. Material

En esta investigación se ha utilizado como matriz de los MMCs, polvo de acero inoxidable austenítico 316L de *Concept Laser GmbH* con la composición química dada en la Tabla 1. Las partículas tienen una morfología esférica, debido a su proceso de fabricación (atomización por gas), con presencia de algunos satélites (Figura 1, arriba) y un rango de tamaño de partícula comprendido entre 25 y 52 micras con un tamaño medio de 27 micras (Figura 1, abajo).

**Tabla 1.** Composición química del acero inoxidable 316L dada por el fabricante (Concept Laser).

| Elemento | Ni    | Cr    | Mo   | Mn   | Si  | V    | Co    | Cu    | C     | S     |
|----------|-------|-------|------|------|-----|------|-------|-------|-------|-------|
| %peso    | 11.55 | 16.53 | 2.00 | 0.54 | 0.7 | 0.06 | 0.062 | 0.029 | 0.015 | 0.007 |

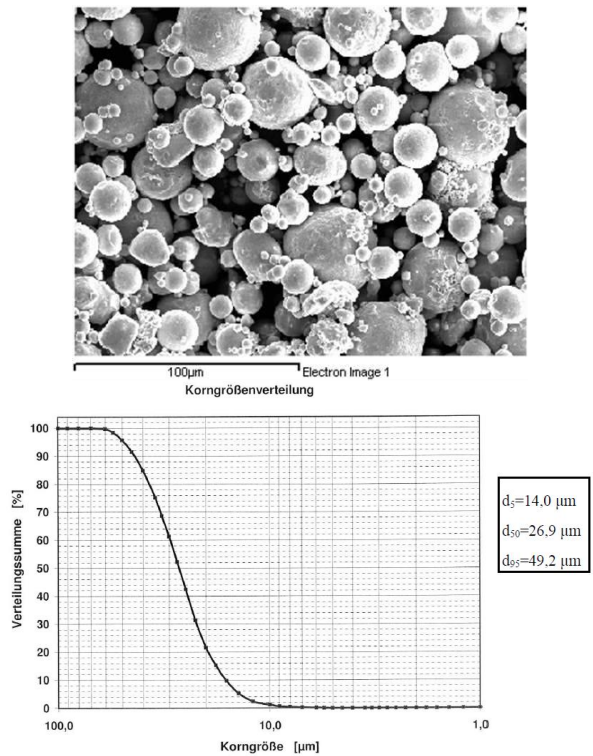

**Figura 1**. Morfología del polvo de acero inoxidable austenítico 316L (arriba) y gráfico de distribución de tamaños de partículas del polvo de acero inoxidable 316L (abajo). **Fuente**: Concept Laser GmbH.

Se utilizó como refuerzo el material cerámico carburo de cromo ($Cr_3C_2$) de la empresa *Flame Spray Technologies* con un tamaño de partícula entre 10 y 45 micras, y una morfología poligonal (Figura 2).

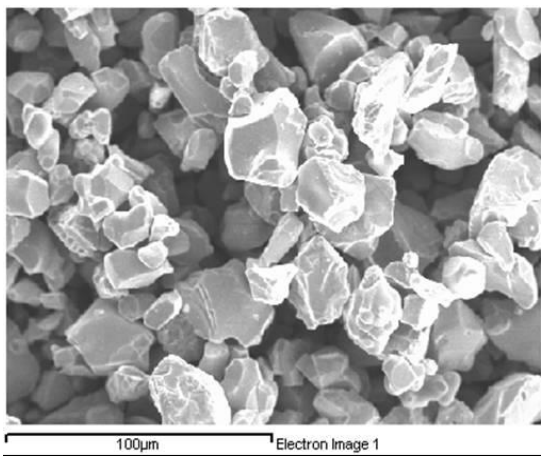

**Figura 2.** Morfología de las partículas cerámicas de $Cr_3C_2$. 500x.

El carburo de cromo o carburo de cromo (II) ($Cr_3C_2$) es un material extremadamente duro de cerámica refractaria, por lo general, se procesa por sinterizado y tiene el aspecto de un polvo gris con estructura cristalina ortorrómbica. Es altamente resistente a la corrosión, y no se oxida, ni siquiera a alta temperatura (1000-1100 °C). El coeficiente de dilatación térmica del carburo de cromo es casi igual al del acero, reduciendo la acumulación de tensiones en la capa límite cuando ambos materiales están unidos o soldados [30].

Para la obtención de los MMCs, se realizó una mezcla convencional del acero 316L y el refuerzo cerámico durante 4 horas en una túrbula vibradora de 30 kg de polvo con un 97, 94 y 91% en peso de 316L y un 3, 6 y 9% en peso de $Cr_3C_2$, respectivamente. Se fabricaron probetas (longitud = 120 mm, diámetro = 14 mm) sin refuerzo y reforzadas para realizar la comparación de las propiedades mecánicas (dureza) y tribológicas (desgaste). Las probetas se fabricaron de modo que el barrido del haz láser se hiciera paralelo (fabricación horizontal) y perpendicular (fabricación vertical) al eje de la probeta (Figura 3b) para evaluar el efecto de la orientación de fabricación sobre las propiedades y optimizar este efecto a la hora de fabricar piezas a gran escala. A estas probetas se les realizó un mecanizado previo a los ensayos de desgaste y así eliminar la influencia de la rugosidad superficial.

### 2.2. Parámetros de procesado para SLM

Para la fabricación de las probetas de ensayo se ha utilizado una máquina de fabricación aditiva mediante fusión selectiva con láser (*Laser Cusing®*) de CONCEPT LASER Modelo M3 (Figura 3a) que se encuentra en el instituto tecnológico AIDIMME (Valencia, España). Este equipo tiene un láser de Nd:YAG con una velocidad nominal de fabricación de 2-5 $cm^3/h$, una precisión nominal de ± 0.1-0.2 mm, espesor de capa de 30 micras y atmósfera de $N_2$ con un contenido de oxígeno residual (% $O_{atm}$) de 0.5%. Los parámetros utilizados se muestran en la Tabla 2. Cabe aclarar que estos parámetros son los dados por el fabricante de la tecnología SLM para el procesado del acero 316L y en este estudio también se utilizaron para los MMCs, ya que se quería analizar si se podían procesar con los mismos parámetros o era necesario optimizarlos y no introducir otras variables adicionales.

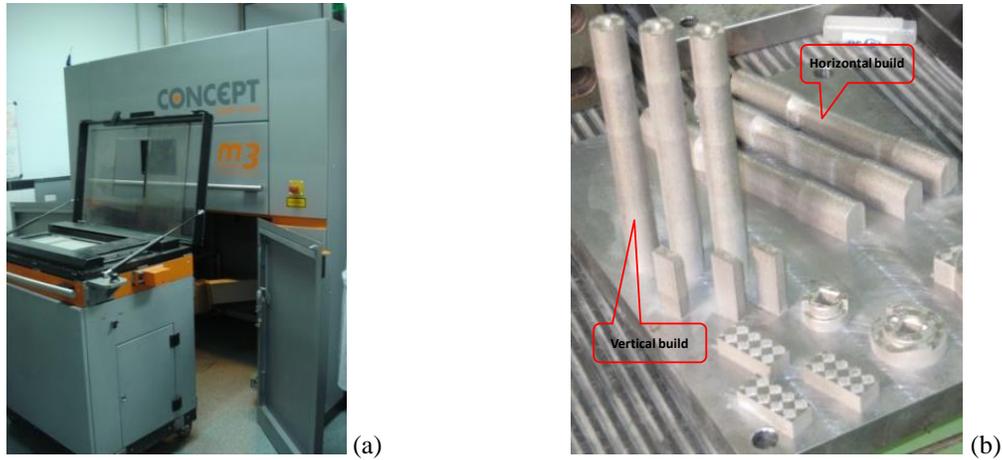

**Figura 3.** a) Máquina CONCEPT LASER modelo M3 de AIDIMME, b) probetas fabricadas en las dos direcciones de fabricación: horizontal y vertical ($L = 120$ mm, $\phi = 14$ mm).

**Tabla 2.** Parámetros de proceso en la tecnología SLM para el procesado de los MMCs.

| | SKIN | | | | | HATCHING | OVERLAPP | | EXPOSURE |
|---|---|---|---|---|---|---|---|---|---|
| | LASER POWER | | | | | | | | |
| Parámetros láser | Potencia (W) | Intensidad (A)* | Velocidad de barrido (mm/s) | Diámetro del foco (mm) | Modo de operación (CW o PULSE) | Orientación | a1 (n x Ø) | a2 (n' x Ø) | Cada n" slice |
| 316L | 100 | 40,1* | 400 | 0.2 | cw | Dámeros perpendiculares | 0.7 | 0.13 | 1 |

| | CONTORNO | | | | | BELICHTUNG |
|---|---|---|---|---|---|---|
| | LASER POWER | | | | | |
| Parámetros láser | Potencia (W) | Intensidad (A)* | Velocidad de barrido (mm/s) | Diámetro del foco (mm) | Modo de operación (CW o PULSE) | Cada n" slice |
| 316L | 100 | 40,1* | 500 | 0.2 | cw | 1 |

| | SOPORTES | | | | | EXPOSURE |
|---|---|---|---|---|---|---|
| | LASER POWER | | | | | |
| Parámetros láser | Potencia (W) | Intensidad (A)* | Velocidad de barrido (mm/s) | Diámetro del foco (mm) | Modo de operación (CW o PULSE) | Cada n" slice |
| 316L | 75 | 36* | 200 | 0.2 | cw | 1 |

*Depende del estado del láser y va en función de la potencia deseada

### 2.3. Caracterización microestructural, mecánica (dureza) y tribológica

Para la caracterización microestructural se obtuvieron micrografías con Microscopía Óptica (OM) en un equipo NIKON ECLIPSE LV100 y SEM (Microscopía Electrónica de Barrido) en un equipo JEOL JSM 6400 de las muestras en estado de pulido y atacadas con el reactivo "Glyceregia" (15 ml HCl + 10 ml Glycerol + 5 ml $HNO_3$) según ASTM E 407-99 [31]. Se realizó el ensayo de dureza Rockwell C en un durómetro Wolpert 660RLD INV-00966 según la norma ASTM E18-05 [32].

El estudio de las propiedades tribológicas se ha realizado mediante ensayos de desgaste tipo "pin on disk" según la norma ASTM G99 – 03 [33] en un tribómetro MT/60/SCM marca Microtest. Los parámetros del ensayo de desgaste fueron los siguientes:

a. **Descripción del Pin**

Tipo: ☒Ball Pin ☐Flat Pin
Material (es): Acero al Cromo
Dureza: 60 HRC
Diámetro [mm]: 6

b. **Descripción de la Muestra**

*Material (es)*: Acero 316L y materiales compuestos con matriz de acero inoxidable austenítico 316L reforzados con 3, 6 y 9 % en peso de $Cr_3C_2$.

*Observaciones*: Estos materiales han sido obtenidos con tecnologías de fabricación aditiva por haz láser (*Laser Cusing*)

c. **Parámetros de Ensayo**

Radio de desgaste [mm]: 3
Carga Normal [N]: 10
Distancia de deslizamiento [m]: 1000
Velocidad de rotación [RPM]: 250
Temperatura [°C]: 25
Humedad relativa [%]: 35
Lubricación: ☐Con ☒Sin

También se analizaron las pistas de desgaste para estudiar los tipos de desgaste involucrados.

## 3. RESULTADOS Y DISCUSIÓN

Con respecto al procesamiento de los MMCs cabe resaltar que las probetas fabricadas con un 9% en peso de partículas cerámicas de $Cr_3C_2$ presentaron una elevada cantidad de grietas, probablemente debido a las tensiones internas generadas por la rápida solidificación y por la interfaz matriz/refuerzo como se observa en la Figura 4. En ambas orientaciones de fabricación se observó el mismo fenómeno. La razón básica por la cual los MMCs son susceptibles al agrietamiento es la existencia de tensiones residuales, las cuales se componen de tensiones térmicas y tensiones de contracción [34, 35]. Si la tensión residual es mayor que la resistencia, se produce la fractura.

Ghosh, S. K. et al [36] evaluaron la densidad de grietas y la resistencia al desgaste de un material compuesto de matriz metálica base Al (Al-MMC) reforzado con partículas de SiC (SiCp) fabricado mediante la tecnología de fabricación aditiva DMLS (*Direct Metal Laser Sintering*). Ellos encontraron que la naturaleza no-uniforme de la distribución de temperatura en la zona afectada durante el calentamiento, fusión y solidificación conduce a un gradiente de temperatura, que eventualmente induce tensiones térmicas en los MMCs. También dilucidaron que las tensiones de contracción en el proceso de sinterización fueron de dos tipos. El primero se originó por la contracción volumétrica del material base desde la curva *liquidus* a la de *solidus*, que se generó principalmente a partir de la transformación de fase. El segundo fue causado por la contracción volumétrica de la curva *solidus* a la temperatura ambiente. A su vez, encontraron grietas a través del SiCp y a lo largo de la interfaz material base/partículas debido a la generación de un gradiente de temperatura entre el material base y el refuerzo.

### 3.1. Microestructura

En la Figura 5 se tienen las micrografías en estado de pulido de todos los materiales procesados (MMCs y el metal base; 316L). Se puede observar una distribución homogénea de las partículas de refuerzo y buena mojabilidad del refuerzo en la matriz. Se presentó porosidad esférica en el metal base (316L) y en los MMCs debido al atrapamiento de gas en el polvo de partida.

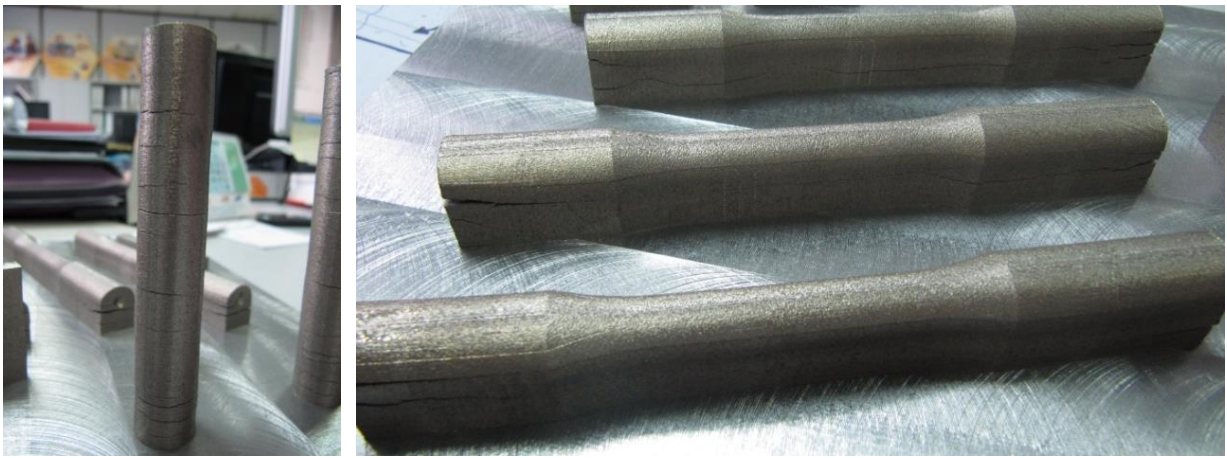

**Figura 4.** Probetas de 316L+9% en peso de $Cr_3C_2$ ($L = 120$ mm, $\phi = 14$ mm).

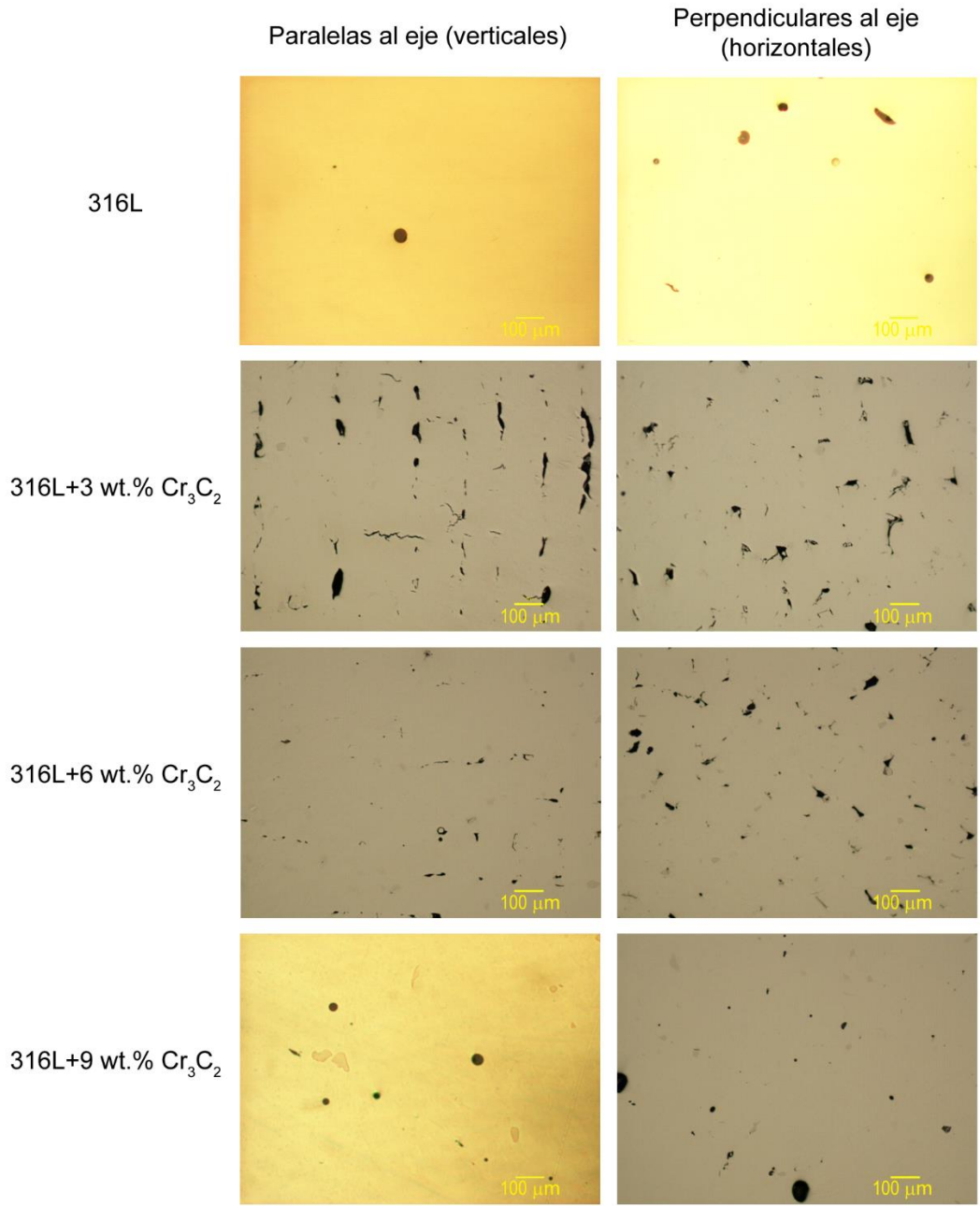

**Figura 5.** Micrografías en estado de pulido para todos los materiales procesados (100x).

La porosidad también es debida a la volatilización de elementos de aleación de bajo punto de fusión durante el procesado de los materiales. Se observó falta de fusión sólo en los MMCs debido a los parámetros de procesado (Figura 6).

En la Figura 7, con las muestras atacadas, se puede ver claramente las zonas fundidas con el láser, revelando la estrategia de fusión láser empleada en el proceso SLM. Se observa el solapamiento de las capas debido a los parámetros del proceso de fabricación (Figura 7 (a) y (b)) y no se ha observado el efecto *balling*, el cual es debido principalmente a una reducción de la mojabilidad. La microestructura de un acero inoxidable 316L y los MMCs procesados por esta técnica es compleja y variada, como cabría esperar en una estructura de solidificación rápida. En la imagen SEM (Figuras 7 (c) y (d)) se observan estructuras columnares paralelas en la dirección de la solidificación (eje Z o plano X-Z) y granos equiaxiales con dendritas muy finas cerca de la interfase del área de fusión en el plano X-Y. En las regiones solapadas se observó el fenómeno de auto-revenido debido a la elevada entrada de calor. No se observaron fenómenos de difusión entre la matriz y el refuerzo (Figuras (e) y (f)).

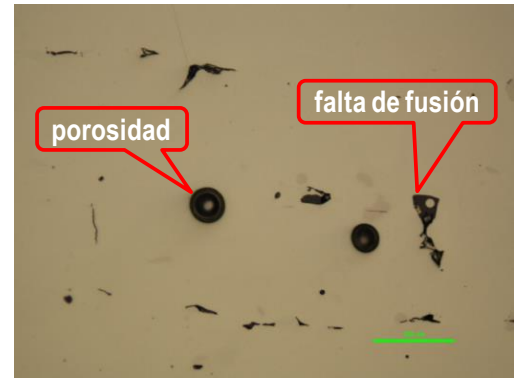

**Figura 6.** Defectos encontrados en el procesado de los MMCs.

Tomando como referencia el material 316L y sus condiciones de procesamiento, la Tabla 3 ofrece una comparación de la porosidad obtenida al añadir el refuerzo cerámico en diferentes porcentajes en peso. Debe mencionarse que, aunque las mediciones se han tomado sólo para una sección (plano X-Y), nos puede dar una idea del orden de magnitud. Teniendo esto en cuenta, se evidencia claramente un aumento de la porosidad a medida que aumenta el porcentaje de refuerzo, especialmente en la dirección vertical. Esto pone de manifiesto la necesidad de optimizar los parámetros de la máquina para cada uno de los MMCs. La mayor densificación es del 99.21% y se obtiene en el material base (316L), orientación horizontal.

**Tabla 3.** Cálculo de la porosidad para los materiales procesados

| Orientación de fabricación | 316L | 316L+3wt.%$Cr_3C_2$ | 316L+6wt.%$Cr_3C_2$ | 316L+9wt.%$Cr_3C_2$ |
|---|---|---|---|---|
| **Horizontal** | 0.79 | 1.25 | 2.44 | 3.52 |
| **Vertical** | 0.85 | 2.76 | 4.38 | 5.69 |

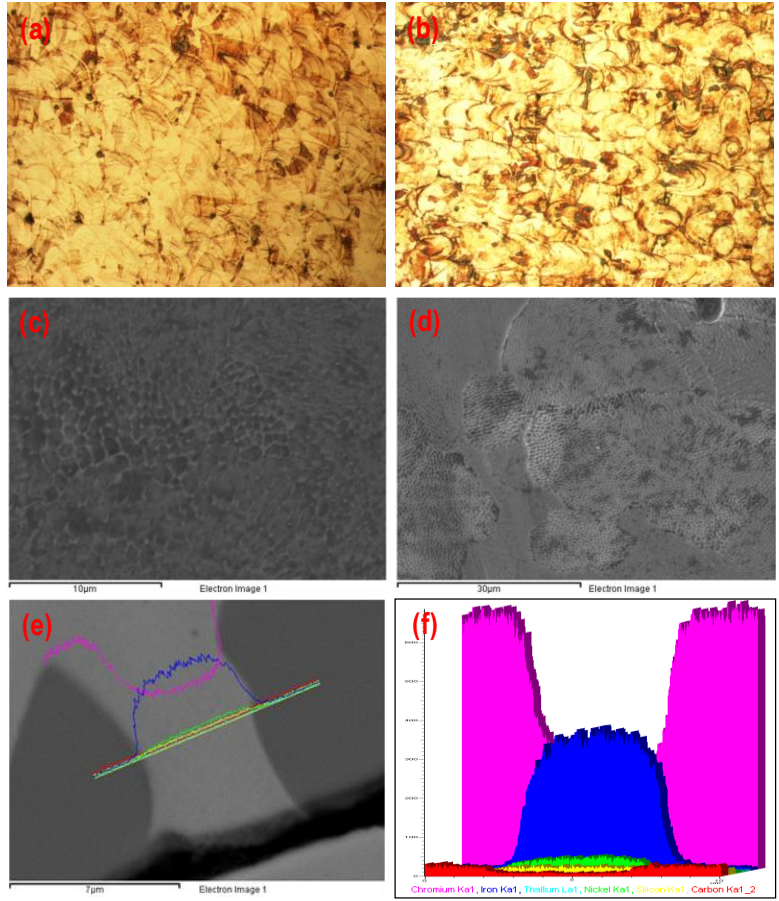

**Figura 7.** Imágenes de microscopía óptica (OM) y SEM atacadas del 316L (a) y (c) horizontal, (b) y (d) vertical. (e) y (f) Análisis en línea para evaluar la interfaz matriz/refuerzo.

### 3.2. Propiedades mecánicas (Dureza)

En la Figura 8 y Tabla 4 tenemos los resultados del ensayo de dureza realizado a todos los materiales procesados. Podemos observar que la dureza aumenta a medida que aumenta el porcentaje de refuerzo hasta el 6% y para el 9% no se aprecia un aumento significativo.

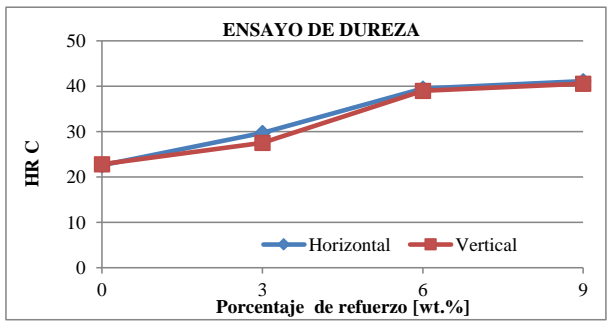

**Figura 8.** Resultados del ensayo de dureza para los materiales procesado por SLM.

**Tabla 4.** Resultados del ensayo de dureza para los materiales procesado por SLM. Desviación típica para siete medidas.

| Porcentaje de refuerzo | Durezas H | Desviación típica | Durezas V | Desviación típica |
|---|---|---|---|---|
| 0 | 22.6 | 0.3 | 22.8 | 0.5 |
| 3 | 29.7 | 0.3 | 27.5 | 0.1 |
| 6 | 39.5 | 0.3 | 39.0 | 0.3 |
| 9 | 41.1 | 0.3 | 40.6 | 0.3 |

### 3.3. Propiedades tribológicas (coeficiente de fricción y tasa de desgaste)

Los resultados del ensayo de desgaste se muestran en la Tabla 5 y en las Figuras 9 y 10 se observa la comparación de los resultados obtenidos.

**Tabla 5.** Resultados del ensayo de desgaste.

| Muestra | Coeficiente de fricción, µ | Tasa de desgaste, W = Δg/P×sd [g/N×m] |
|---|---|---|
| 316L_H | 0.64 | $3.80 \times 10^{-6}$ |
| 316L_V | 0.63 | $3.60 \times 10^{-6}$ |
| 316L+3%wt. $Cr_3C_2$_H | 0.67 | $2.63 \times 10^{-6}$ |
| 316L+3%wt. $Cr_3C_2$_V | 0.68 | $2.07 \times 10^{-6}$ |
| 316L+6%wt. $Cr_3C_2$_H | 0.67 | $8.00 \times 10^{-7}$ |
| 316L+6%wt. $Cr_3C_2$_V | 0.65 | $7.00 \times 10^{-7}$ |
| 316L+9%wt. $Cr_3C_2$_H | 0.69 | $1.00 \times 10^{-7}$ |
| 316L+9%wt. $Cr_3C_2$_V | 0.69 | $1.00 \times 10^{-7}$ |

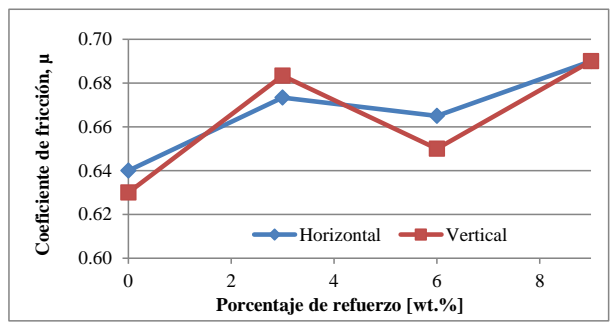

**Figura 9.** Coeficiente de fricción de los MMCs y el metal base 316L en las dos direcciones de fabricación. Desviación típica de 0.01.

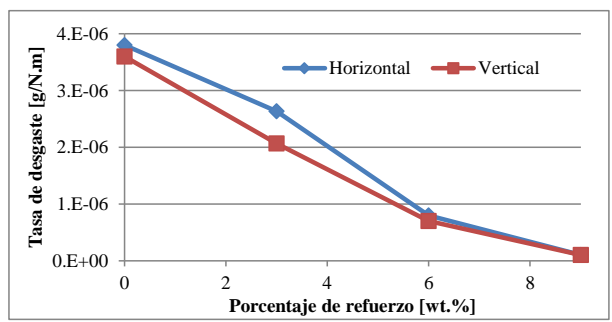

**Figura 10.** Tasa de desgaste de los MMCs y el metal base 316L en las dos direcciones de fabricación.

Se tomaron tres medidas para cada punto. El coeficiente de fricción está entre 0.63 y 0.69, y no se observa un patrón de comportamiento. En la Figura 10 se hace evidente que el acero inoxidable 316L presenta una mayor pérdida material en comparación con los MMCs, mientras que la adición de partículas de $Cr_3C_2$ a la matriz de acero inoxidable fue muy eficaz para reducir la tasa de desgaste. El mejor comportamiento se obtuvo para el material 316L + 9% $Cr_3C_2$ a pesar de que estas muestras tienen el mayor porcentaje de porosidad, es decir, que la porosidad no juega un papel dominante y sí lo hacen las partículas de refuerzo, que han aumentado la dureza de los materiales compuestos de acero inoxidable, como se indica en la Figura 8. Además, la distribución homogénea de los refuerzos finos es favorable para la mejora de la resistencia al desgaste. Sulima, I. [37] encontró el mismo comportamiento para un MMC con matriz de acero 316L y refuerzo de $TiB_2$ pero procesado por SPS (*Spark Plasma Sintering*).

Realizando una comparación entre las dos direcciones de procesado para cada uno de los materiales, no se observa una variación significativa en los valores de coeficiente de fricción y tasa de desgaste.

En la Figura 11 se observan las pistas de desgaste de los materiales procesados. Para todas las pistas de desgaste se observan marcas de desgaste a lo largo de la dirección de deslizamiento, lo que es típico para una superficie de desgaste junto con la presencia de surcos o ranuras de arado. También se presentan signos de grietas de delaminación. Sin embargo, no se observa una delaminación significativa.

Las pistas de desgaste se cubren con una película de óxido de un color blanco, microscópicamente suave y una apariencia superficial agrietada. En algunas regiones, la película de óxido se elimina para formar los residuos de desgaste, dejando atrás una superficie oscura y áspera. Esto es típico del desgaste oxidativo que es responsable de un desgaste leve en muchos materiales. La oxidación de las superficies de deslizamiento es el resultado de calentamiento por fricción en atmósfera ambiente. Para que el desgaste oxidativo llegue a ser el desgaste predominante, se debe formar una película de óxido de espesor suficiente y la superficie y sub-superficie del material debe tener suficiente integridad para apoyar la película de óxido durante el deslizamiento. En las Figuras 11 (a) y (b) se observa la película de óxido acumulado en la pista de desgaste del acero 316L. Bajo condiciones de deslizamiento repetidas, la película de óxido está a punto de desprenderse para formar partículas de desgaste. Por otra parte, la huella de desgaste tiene diferentes morfologías. La pista de desgaste es rugosa/áspera y tiene un aspecto metálico. Muchas ranuras, cráteres y grietas se pueden encontrar en las pistas de desgaste en todas las muestras. Se puede evidenciar la fractura de material alrededor de los poros preexistentes en la pista de desgaste. En los bordes de la pista de desgaste se puede ver claramente la formación de grietas, la propagación y la fractura de material alrededor de los poros, por lo tanto, el principal mecanismo de desgaste es la fractura debido a la iniciación y propagación de fisuras procedentes de los poros preexistentes en la pista de desgaste.

**Figura 11.** Pistas de desgaste de los materiales procesados: 316L en la dirección horizontal (a) y vertical (b), 316L+3wt.%Cr$_3$C$_2$ en la dirección horizontal (c) y vertical (d), 316L+6wt.%Cr$_3$C$_2$ en la dirección horizontal (e) y vertical (f), y 316L+9wt.%Cr$_3$C$_2$ en la dirección horizontal (g) y vertical (h).

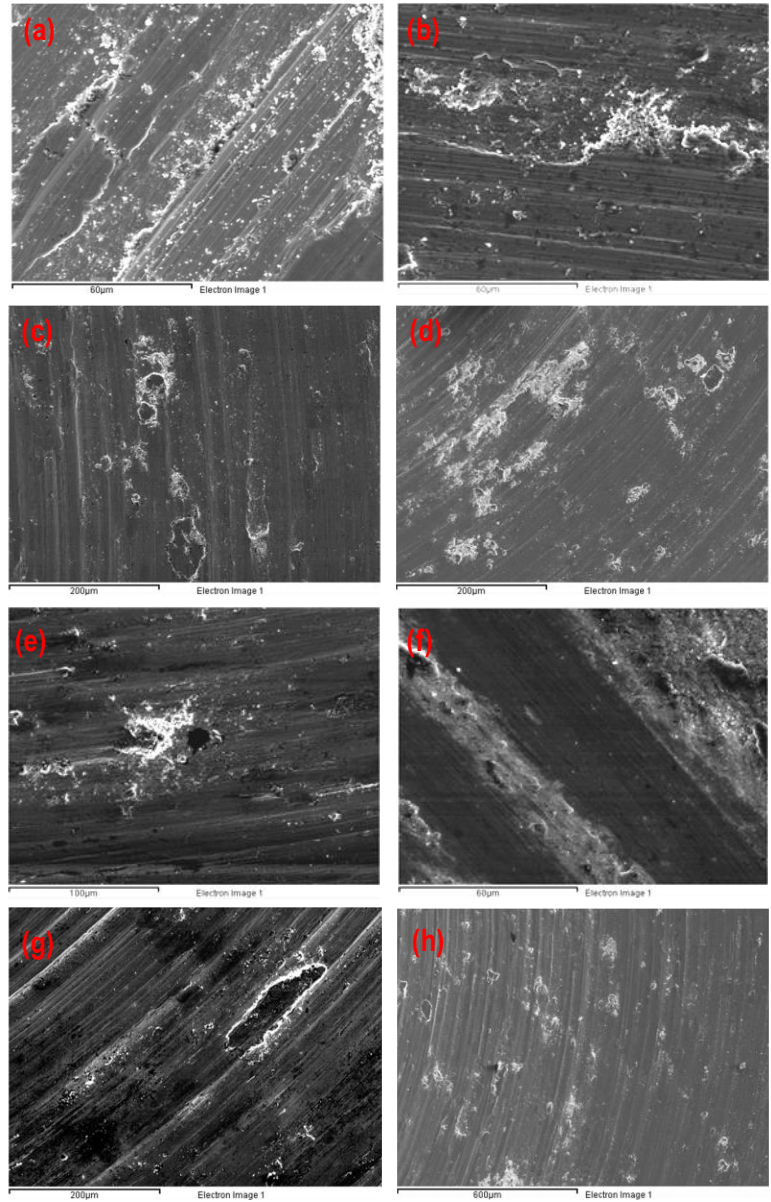

Los surcos de arado fueron disminuyendo a medida que aumentaba el porcentaje de refuerzo. En la Figura 11(g) se puede observar la presencia de grietas de delaminación y una capa parcialmente delaminada junto con algunas partículas de óxido.

## 4. CONCLUSIONES

De esta investigación se obtuvieron las siguientes conclusiones:

**4.1** Fue posible procesar de forma adecuada MMCs con tecnologías de fabricación aditiva por haz láser (SLM).

4.2 Es necesario optimizar los parámetros de la máquina para cada % de refuerzo y así obtener una densificación mayor.

4.3 En la microestructura se observa el solapamiento de las capas debido a los parámetros introducidos al proceso de fabricación. En general, predominan las estructuras columnares paralelas a la dirección de solidificación (eje Z) y granos equiaxiales cerca de la interfase del área de fusión (plano X-Y).

4.4 La dureza aumenta con el porcentaje de refuerzo hasta el 6%, y para el 9% no se aprecia un aumento significativo, siendo muy similar al valor del 6%. Los valores de dureza para las dos orientaciones de fabricación son muy similares.

4.5 No se observa un patrón de comportamiento en el coeficiente de fricción a medida que aumenta el porcentaje de refuerzo mientras que la tasa de desgaste si lo hace con una disminución sustancial. Sin embargo, no hay una variación significativa en los valores de coeficiente de fricción y tasa de desgaste cuando se compara para un mismo material en las dos direcciones de procesado

4.6 Hay una predominancia de desgaste oxidativo.

## 5. REFERENCIAS